# A Framework for Predicting Impactability of Healthcare Interventions Using Machine Learning Methods, Administrative Claims, Sociodemographic and App Generated Data

Heather Mattie PhD,[1] Patrick Reidy MPH,[2] Patrik Bachtiger MD, MSc,[1]
Emily Lindemer PhD,[2] Mohammad Jouni MS,[2]  Trishan Panch MD, MPH[2]

1. Harvard T.H. Chan School of Public Health, Boston, Massachusetts
2. Wellframe Inc., Boston, Massachusetts

April 2019

# Table of Contents

## Summary

'Low risk' populations typically receive less intervention from health insurance providers' preventative programs. However, focusing on 'high risk' populations misses early stage chronic disease, which offers some of the most compelling opportunities for cost-effective interventions that improve health outcomes. Digital care management programs can reduce healthcare costs, improve the quality of care and be implemented on a scale that covers the spectrum of patient risk. However, it is not clear how to target patients who are most likely to benefit from these programs ex-ante, a further shortcoming of current 'risk score' based approaches.This study focuses on defining impactability (as a dollar value), by identifying those patients most likely to benefit from technology enabled care management, delivered through a digital health platform, including a mobile app and clinician web dashboard. Anonymized insurance claims data were used from a commercially insured, low-risk population across several U.S. states and combined with inferred sociodemographic data and data derived from the patient-held mobile application itself. Our approach involves the creation of two models and the comparative analysis of the methodologies and performances therein. We first train a cost prediction model to calculate the differences in predicted (without intervention) versus actual (with onboarding onto digital health platform) healthcare expenditure for patients ($N = 1{,}242$). This enables the classification of impactability if differences in predicted versus actual costs meet a predetermined threshold. A random forest machine learning model was then trained to accurately categorize new patients as impactable versus not impactable, reaching an overall accuracy of 71.9%. We then modify these parameters through grid search to define the parameters that deliver optimal performance. This demonstrates the potential to successfully target, based on impactability, lower risk members in the payor population with a digital health intervention. However, in the context of low-cost digital health interventions, wide inclusion is possible. Instead, an impactability-based machine learning model, such as the one described here, is able to maximise efficient spending of non-operational costs e.g. patient recruitment. A roadmap is proposed to iteratively improve the performance of the model. As the number of newly onboarded patients and length of use continues to increase, the accuracy of predicting impactability will improve commensurately as more advanced machine learning techniques such as deep learning become relevant. This approach is generalizable to analyzing the impactability of any intervention and is a key component of realising closed loop feedback systems for continuous improvement in healthcare.

## Introduction

Efficient use of resources is a priority for healthcare systems worldwide. An ageing, multimorbid population is demanding more health services, placing pressure on systems already struggling with and a lack of interconnectivity between providers, low productivity, poor adaptability[1] and lack of patient engagement. Care management programs can ameliorate some of these pressures by enabling effective coordination of care towards preventing avoidable need for healthcare utilization.[2]

Previous studies of care management interventions assert that savings can only be achieved by targeting those with the highest number of chronic diseases, who also have the highest healthcare costs.[3,4] However, selectively targeting interventions to high-cost, high-need (HCHN) patients assumes that those with the highest health expenditures are reservoirs of untapped savings. Across a broad variety of chronic care management programs remunerated by Medicaid targeting HCHN patients, the anticipated savings have fallen short. On average, across 34 programs, there was no effect on hospital admissions or regular Medicare expenditures.[5]

The concept of impactability reframes how to allocate healthcare resources by matching individuals to interventions based on their ability to benefit from these. 'Benefit' in the analysis laid out here is most easily measured by identifying a reduction in healthcare costs (in $US), which acts as a proxy for improvements in health outcomes.[7] However,

attainments other than cost-saving, such as clinical outcomes or patient reported outcome measures (PROMS), can also be applied to a measure of impactability. As a relatively new approach to resource allocation, work by DuBard and Jackson generated an 'impactability score' that uses administrative data to predict achievable savings in a care management population in North Carolina.[8] Aligning with the principles of impactability, Hawkins et al developed a propensity-to-succeed (PTS) score to prioritize individuals most likely to benefit from a care coordination intervention.[9] Put simply, impactability reframes how to allocate healthcare interventions by identifying patients to target for the greatest return on investment, thereby acknowledging the reality that patients with the highest healthcare cost may not overlap with being the most impactable.

The richness of the data collected by digital health management solutions makes these an opportune tool for exploring impactability. Adoption of digital tools is increasing across healthcare, ranging from wearable devices to mobile-enabled disease and care management applications. One of the benefits of digital health solutions as opposed to in person consultations is that they have the potential to drastically reduce the cost of care delivery, meaning that care can be offered to more people if that care can be safely and successfully encapsulated in a digital health intervention. Digital health tools have added policy relevance as they can help realise a paradigm shift from a fee-for-service to a value-based care model, namely by offering improved patient outcomes with the added virtue of being cost-effective.[6,10] Care management has been particularly successful in implementing digital health tools.[11–13] However, in the face of increasingly restricted budgets, whom to target with these interventions is a pertinent question for risk bearing organisations.

For care management, the relatively low start-up and maintenance cost of digital health solutions already enables the inclusion of more members. It is also important to acknowledge that digital health solutions clearly offer benefits to patients and providers that go beyond cost-savings. In the context of digital care management at least, impactability-based allocation should not be the rationale for offsetting the cost of setup and day-to-day upkeep of the technology. These interventions can, however, have significant external costs, namely the marketing campaigns needed to recruit patients. In the context of a widely available digital health care management tool, care management teams can use an impactability-based approach to target marketing campaigns, thus maximising recruitment of those patients most likely to benefit (and subsequently offset the cost of marketing).

It is through recent technological advances that refinement of an impactability based approach has become possible. Most digital health tools share a vital strength: the ability to collect data. Enabled by the combination of ever-larger datasets and increasing computing power, machine learning (ML) methods are now able to effectively 'learn' from these data, highlighting patterns and in many instances exhibiting judgement superior to that of humans.[14] As a branch of artificial intelligence, ML methods transform the inputs of an algorithm into outputs using statistical, data-driven rules that can be automatically derived from a large set of examples. A number of successes include aiding diagnosis of sight-threatening eye disease[15], predicting risk of hospital readmission[16] and predicting onset of type 2 diabetes.[17] Across other industries, ML techniques are being applied in many sectors including law[18], finance[19] and urban planning[20]. However, ML techniques remain largely untested in the domain of impactability-based resource allocation, having previously relied on more classical, less powerful prediction models.[8,9] This highlights an exciting new frontier for judicious allocation of finite resources that aligns itself well with the principles of value-based healthcare.

This study demonstrates an example of how a combination of ML methods can be used to identify the features of patients whose costs significantly declined once onboarded onto a digital health intervention i.e. highlighting the features that constitute impactability. Such an approach allows a subsequent ML algorithm to predict a new patient's impactability, thereby offering a tool to assist in decision-making around allocating the intervention. We used data from a commercially insured population, combining insurance claims, sociodemographic and patient-generated data (from the use of the patient-held mobile app at the core of the digital health intervention described below).

The study population was onboarded onto a digital health tool created by the team at Wellframe Inc, a Boston based digital health company. The Wellframe intervention consists of a of a mobile-enabled care management platform, based around an app that includes a customized, interactive, daily health checklist displayed to patients on their smartphone or tablet. Responses and interactions from users are collected as actionable day-to-day-health information. The app serves to provide a sustained and supportive portal of communication between patients and their care teams. A built-in concierge service helps patients pursue their health goals, get their benefits questions answered and navigate the healthcare system — all through a single, direct channel of communication. The population in this study was onboarded onto the Wellframe app as part of a health coaching programme, both patients and clinical teams received training on how to use all of its features.

**Data**

The majority of the data used came from individual health care claims from June 10, 2015 through May 26, 2018, with other variables created using national databases.[21] Unique patient ID, the date of the visit, type of the visit (inpatient or outpatient), amount paid by the provider for the visit, patient gender (male or female), patient date of birth, patient zip code, and diagnosis(es) were recorded for each claim. Age was calculated as the number of years between the date of birth and December 31, 2018. Education level, employment status, income and poverty status were inferred from the patient's zip code. Education level variables included the percent of individuals living in the same zip code as the patient with an education level of less than high school, high school, some college, an associate's degree, a bachelor's degree, and a graduate degree. Employment status included the percent of individuals living in the same zip code as the patient who were in the labor force and employed, in the labor force and unemployed, and not in the labor force. Income level included two variables: the total income and average income for all individuals living in the same zip code as the patient. Poverty status was represented as the percent of individuals living in the same zip code as the patient living below the 138% Federal Poverty Level.

Developing the impactability model discussed in this paper required integration of multi-modal data and use of this data for model training and testing. This was achieved using the infrastructure shown in figure 1, built using Google Cloud Platform (GCP).

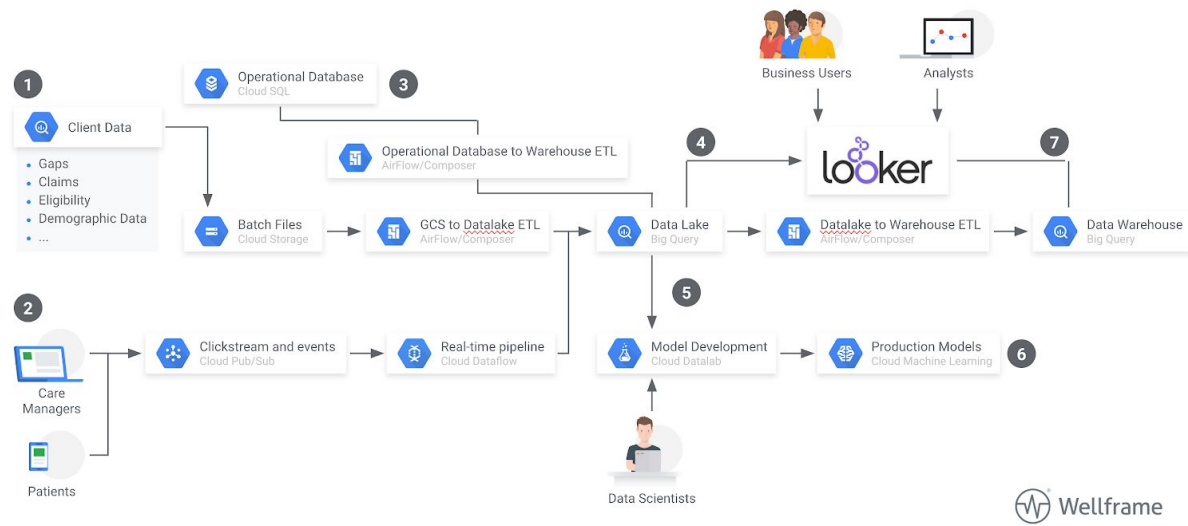

Figure 1. Airflow is a data orchestration service able to ingest client provided claims information and merge it with transactional data generated by the Wellframe platform. The amalgamation of these two data sets was stored into a centralized data warehouse: BigQuery. As a serverless, managed data warehouse, BigQuery allows the deployment of a centralized data warehousing solution without the need to maintain large computing clusters. As a data footprint grows, this can reduce operational cost while giving data scientists flexibility in how they interact with the collected data. Once in BigQuery, a collaboration of data science teams is able to build models by leveraging Cloud datalab and Jupyter Notebooks. They can also access cutting edge tools like Keras, to build deep neural nets and train models using Cloud TPUs (Cloud Tensor Processing units). Once a model is trained, Cloud ML Engine can deploy it into production and make it available via an application programming interface (API). This API can then be utilized by applications

(mobile apps or web apps) to predict a client's propensity to benefit. All operational data generated from the interaction of users with the platform is stored in Cloud SQL database. This data is replicated at regular interval into a data warehouse. The fast availability of this data enables validation that the recommendations offered by the deployed models are actually yielding the intended goals. The data collected can be exposed via Looker to enable business users to access the results of a predictive model. Looker allows businesses to track key conversion metrics that can help assess the impact of the recommendations, without requiring expertise in data.

For each patient, claims were combined by month. The total cost for the month was the sum of the amount paid for each claim that month. Any month a patient did not have a claim the total cost was set to 0 dollars. The rolling cost, the moving average cost of the last 6 months, was also calculated in an effort to smooth out short-term fluctuations and highlight longer-term trends. Total number of inpatient visits per month were calculated from the type of visit reported in the claim. The total number of unique diagnoses was calculated from all claims in a given month.

Finally, the data were split into pre (time before onboarding e.g. before accessing Wellframe) and post (time since onboarding) time periods. The pre time period included monthly data for all visits before onboarding to Wellframe. The post period included monthly data for all visits after onboarding to Wellframe. The month a patient onboarded was designated as $T$ = 0, and the range of values in this data set were $T$ = [-35, 21]. Only patients who had onboarded were used in the analysis, resulting in a sample size of N = 1,280.

## Methods

With the overarching goal of predicting which patients would benefit the most from onboarding, our targeting process includes building two different types of models: a cost prediction model and an impactability prediction model. The cost prediction model is used to predict the monthly cost after time $T$ = 0 for patients had they not onboarded. This can be thought of as estimating the counterfactual cost for these patients who did onboard. After predicting these counterfactual costs, the difference in actual cost and predicted cost is calculated for each patient, an impactable threshold is chosen, patients are labeled as impactable or not, and the impactability model is used to classify patients as impactable or not.

Since the impactability prediction model depends on predictions made from the cost prediction model, the cost prediction model must be built first. We included all individuals who had at least 12 months of claims before onboarding, i.e. only those with a minimum $T \le -12$. This cutoff was chosen for two reasons. The first being only a small proportion of patients had claims information from more than 12 months before onboarding. The second reason was to ensure enough prior information was available for each patient to predict future monthly costs. This slightly decreased the sample size to N = 1,242. See Table 1 for sample descriptive statistics. We then fit several models to predict the rolling cost and the natural log of the rolling cost for months $T = -2$ and $T = -1$ using months $T = -12$ to $T = -3$. We used only the months before onboarding to fit our cost prediction models since our goal is to estimate monthly cost had someone not onboarded. Due to the longitudinal nature of our data, we fit different mixed effects models to account for the repeated measures for each individual. For each type of outcome, rolling total and log rolling total, we fit four models: a random intercept model with time as the only predictor, a random intercept model with all covariates described above as predictors, a random intercept and random slope model with time as the only predictor, and a random intercept and random slope model with all predictors. This resulted in a total of eight models. To judge which model performed the best, we fit each model, predicted the cost for months $T = -2$ and $T = -1$ for each patient, and then calculated the root mean square error (RMSE) and mean absolute deviation (MAD) between predicted cost and actual cost.

Once the best cost prediction model was defined, cost for months $T = 0$, 1 and 2 were predicted using all months before onboarding for each patient. These predictions are used as an estimate of what the costs would have been had the patient not onboarded. The average residual between the predicted and actual cost for the three months was calculated for each patient and patients were then sorted by their residual value. Larger negative residuals indicate predicted costs were higher than actual costs and larger positive residuals indicate predicted costs were lower than actual costs. Patients were then labeled as "impactable" if their residual was below a certain threshold, $h$, and labeled as not impactable if their residual was above this threshold.

We then split the data into training (80%, n = 993) and testing (20%, n = 249) sets. Due to the relatively low sample size we did not use a validation set for this model building step. We then fit several random forest (RF) and logistic regression (LR) models to predict if a patient is impactable or not using their gender, age, total cost for all claims before onboarding, total number of inpatient visits before onboarding, total number of unique diagnoses before onboarding, and all zip code level variables. For both RF and LR methods we fit three different models; one that did not correct for class imbalance, one that corrected for class imbalance through weighting,[22] and one that corrected for class imbalance through the synthetic minority over-sampling technique (SMOTE).[23] The performance of each model was measured by the number of false positives and false negatives for different thresholds of the impactable threshold, $h$, ranging from the 10th percentile to the 90th percentile. In addition to the threshold $h$, a classification threshold $c$ was allowed to vary between 0.5 and 0.9. When $c = 0.5$, a patient is classified as not impactable if the predicted probability of that patient not being impactable is above 50%. As $c$ increases, the model must be more certain a patient is not impactable before classifying them as not impactable.

## Results

The best cost prediction model was deemed the model with the smallest RMSE and MAD. For this data set, this was the random intercept and random slope model with time trend as the only predictor for rolling cost. This model produced an RMSE equal to 1340.49 and MAD equal to 93.80. This cost prediction model produced residuals ranging from -$462 (10th percentile) to $382 (90th percentile) for this population.

The number of false negatives for each RF and LR model and every value of $h$ and $c$ are presented in Figures 2-7. The threshold $h$ in terms of percentile of patients as well as the dollar amount equivalent are presented on the x-axis. The random forest models, models 1, 2 and 3, show monotonic curves that are fairly smooth. As the impactability threshold $h$ increases, the number of false negatives increases for these models. This is not surprising; as $h$ increases, the number of individuals labeled as impactable increases and thus the number of possibilities in classifying an impactable patient as not impactable also increases. Inversely, as the classification threshold $c$ increases, the number of false negatives decreases. This is also intuitive: as the threshold for classifying an individual as not impactable increases, our model must be more confident in it's classification. As a result, more patients will be classified as impactable, resulting in fewer false negatives.

The feature importance values were also recorded for each impactable threshold $h$ for models 1, 2 and 3 to gain insight into which variables were most predictive of impactability. For all three models, across all thresholds, the number of inpatient visits in the pre-period, the total cost in the pre-period, a patient's age, and the number of conditions diagnosed in the pre-period were the most predictive of impactability. Interestingly, the feature importance of gender was higher for values of $h$ that indicate a higher class imbalance. Specifically, gender was more predictive of impactability for low and high values of $h$, but lower for middle values of $h$, around 0.5. This pattern was inverted for all other variables with low predictive power.

The logistic regression models, models 4, 5 and 6, are not monotonic and much less smooth compared to the random forest models. For some combinations of $h$ and $c$ the number of false negatives plummets to 0 or spikes to a maximum point. A value of 0 or close to 0 indicates the model is classifying every patient, or nearly every patient, as impactable. This results in all impactable patients receiving care and no false negatives. A value near the maximum number possible given $h$ and $c$ indicates the model is classifying everyone, or nearly everyone, as not impactable. In this case all or nearly all impactable individuals do not receive care and the number of false negatives increases dramatically.

Varying the thresholds $h$ and $c$ highlight the tradeoff in the number of FPs and FNs. While it is true that deciding to target all patients for on-boarding would eliminate the number of FNs, this may not be feasible for the resources at hand. Depending on the size of the population and resources needed, recruitment costs and effort could be prohibitive and more of a balance in the number of FNs and FPs may be more realistic. With this in mind, the thresholds that result in the best tradeoff for each model, with corresponding sensitivity, specificity, number of false positives and false negatives, and total number of misclassified patients are presented in Table 2. All three RF models limit FPs more than FNs while all LR models limit FNs more than FPs.

| Measured Variables | Sample, N = 1,242 |
|---|---|
| Gender, n (%)<br>    Female<br>    Male | <br>864 (69.6)<br>378 (30.4) |
| Age, M (SD) | 44.8 (11.1) |
| **Inferred Variables** | |
| Education level percent, Med (IQR)<br>    <9th grade<br>    9-12th grade<br>    High school diploma<br>    Some college<br>    Associates degree<br>    Bachelor's degree<br>    Graduate degree | <br>2.2 (3.1)<br>4.1 (3.4)<br>26.0 (12.4)<br>23.4 (5.2)<br>9.6 (3.4)<br>21.1 (13.6)<br>9.5 (9.4) |
| Average income, Med (IQR) | $59,780 ($24,660) |
| Percent insured, Med (IQR) | 92.2 (6.0) |
| In labor force, Med (IQR)<br>    Percent employed<br>    Percent unemployed | <br>69.5 (8.5)<br>2.4 (1.7) |
| Percent not in labor force, M (SD) | 27.8 (7.4) |
| Percent below poverty line, Med (IQR) | 15.5 (7.2) |
| Number of inpatient visits in pre period, Med (IQR) | 8 (28) |
| Number of diagnoses in pre period, Med (IQR) | 14 (13) |
| Total cost in pre period, Med (IQR) | $2,872.80 ($9,365.70) |

Table 1. Measured and Inferred Sample characteristics.

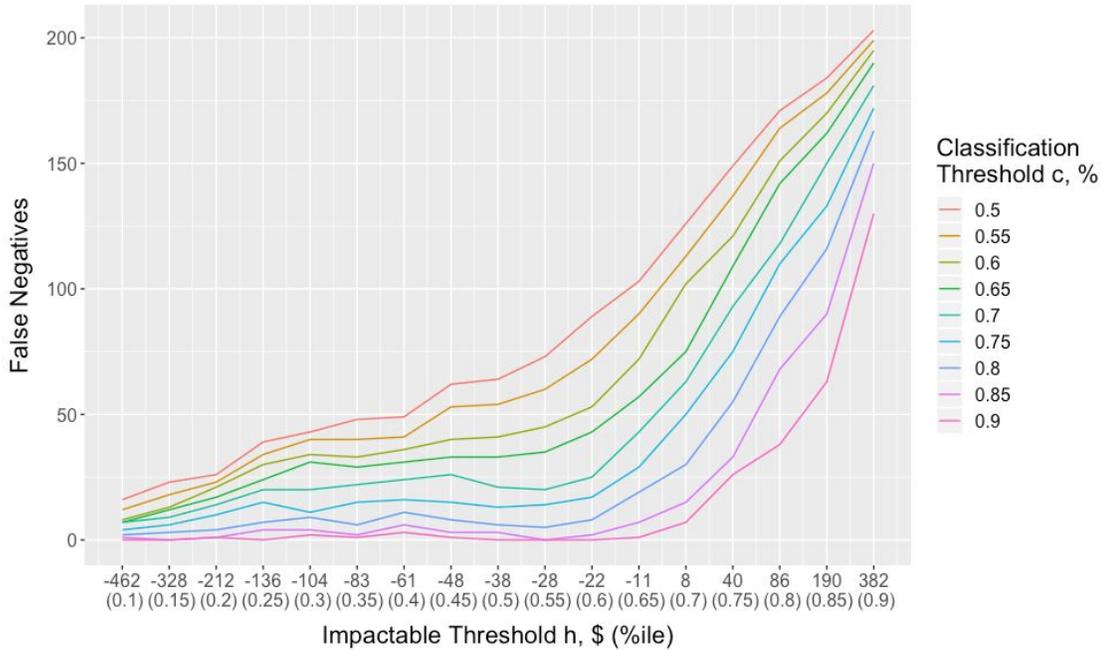

Figure 2. Number of false negatives as a function of impactable threshold *h* and classification threshold *c* for model 1; Random forest without correction for class imbalance. The integers on the x-axis represent the residual dollar amount threshold for labeling a patient as impactable. The decimals in parentheses on the x-axis represent the percentile of the patients in the test set that were labeled as impactable.

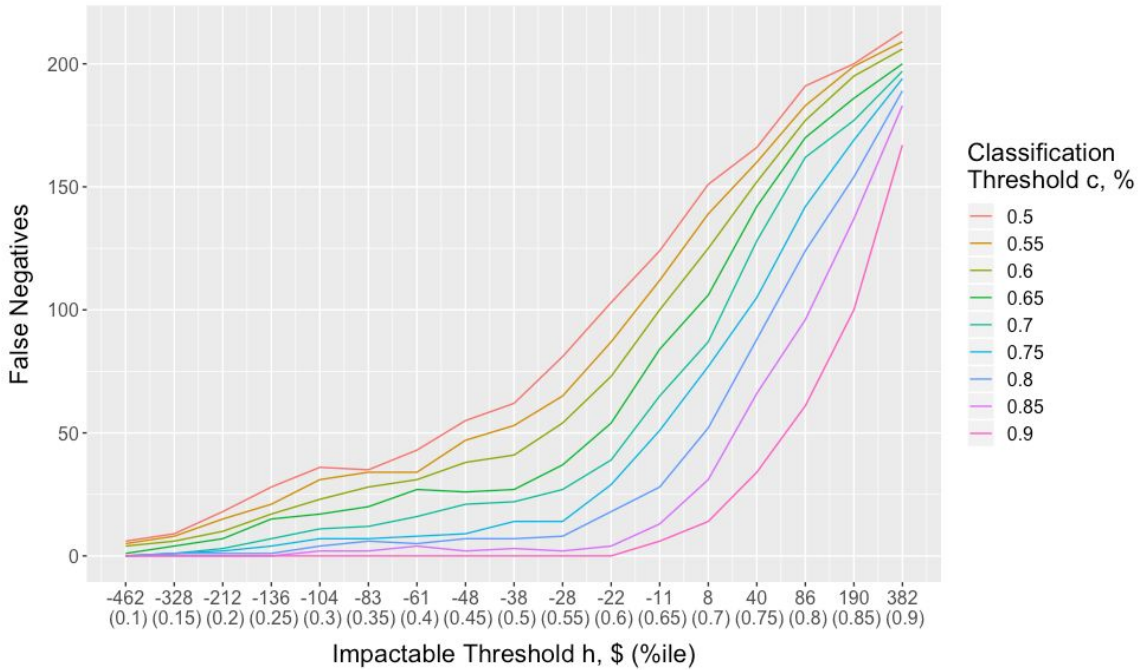

Figure 3. Number of false negatives as a function of impactable threshold *h* and classification threshold *c* for model 2; Random forest with correction for class imbalance using weighting.

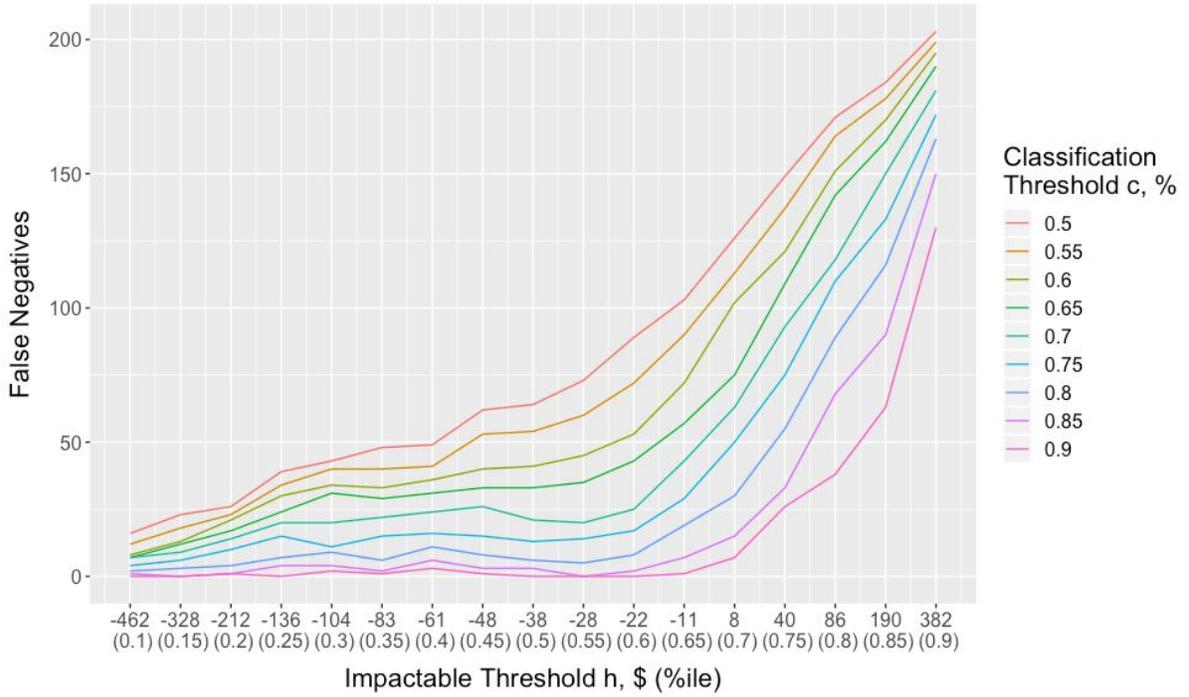

Figure 4. Number of false negatives as a function of impactable threshold *h* and classification threshold *c* for model 3; Random forest with correction for class imbalance using SMOTE.

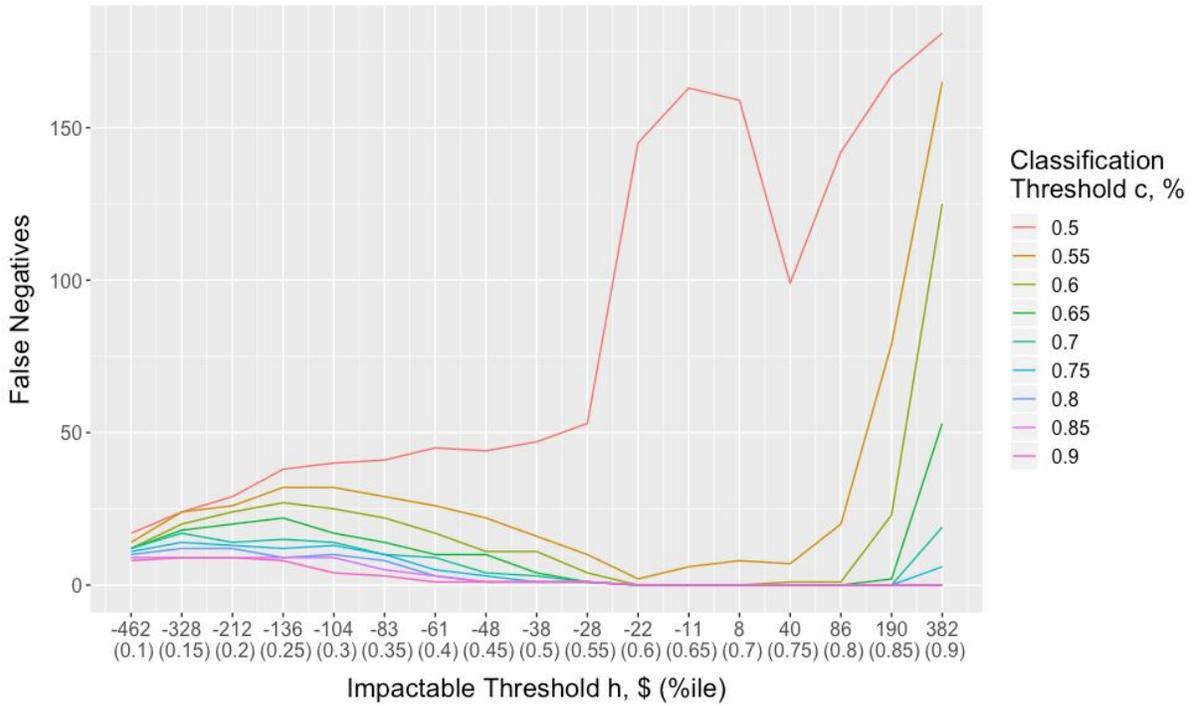

Figure 5. Number of false negatives as a function of impactable threshold *h* and classification threshold *c* for model 4; Logistic regression without correction for class imbalance.

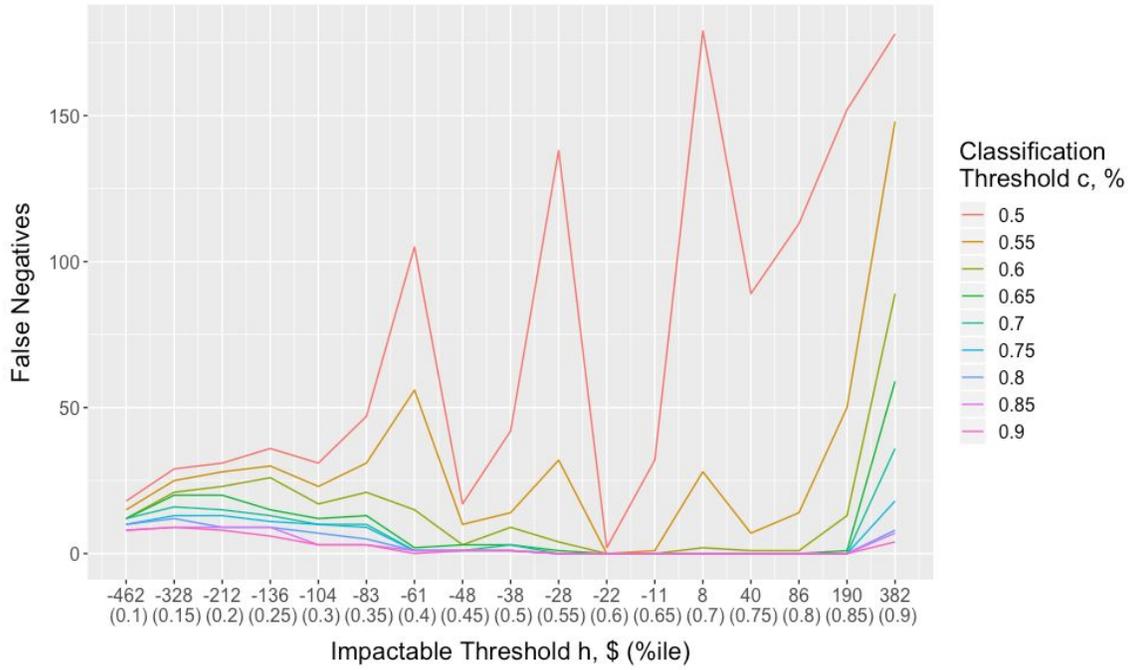

Figure 6. Number of false negatives as a function of impactable threshold *h* and classification threshold *c* for model 5; Logistic regression with correction for class imbalance using weighting.

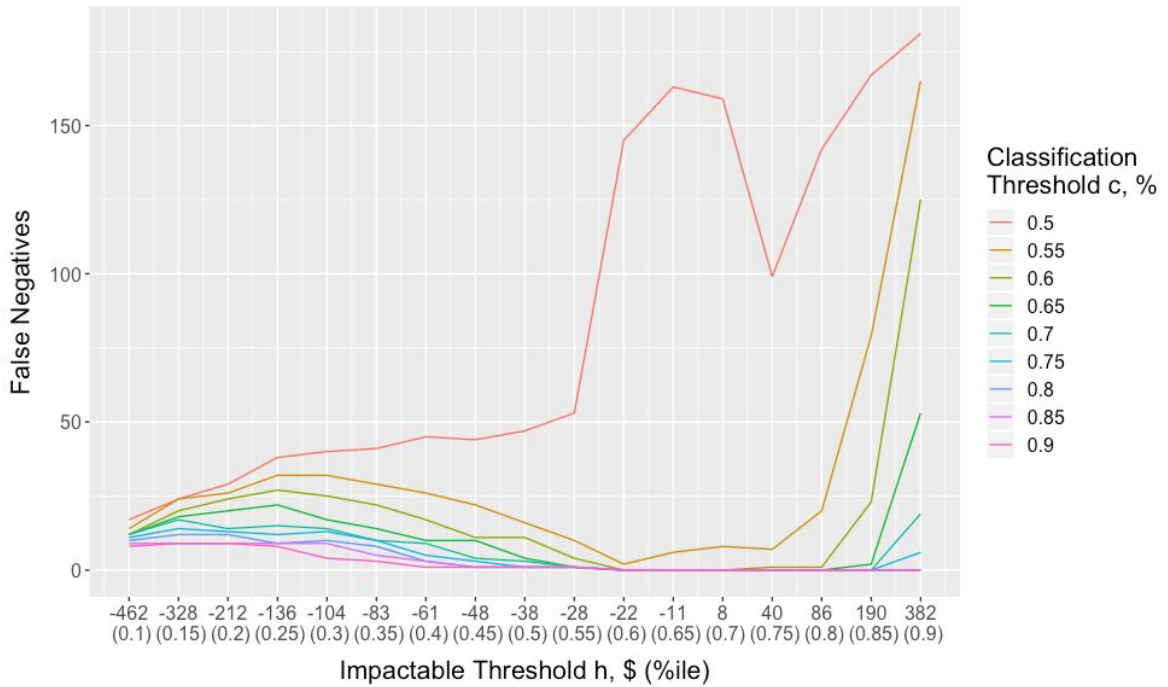

Figure 7. Number of false negatives as a function of impactable threshold *h* and classification threshold *c* for model 6; Logistic regression with correction for class imbalance using SMOTE.

| Model | Impactable threshold *h* | Classification threshold *c* | Sensitivity | Specificity | FP | FN | Total misclassified |
|-------|--------------------------|------------------------------|-------------|-------------|-----|-----|---------------------|
| 1 | -$22 (0.6) | 0.5 | 0.39 | 0.72 | 30 | 89 | 119 |
| 2 | -$22 (0.6) | 0.65 | 0.63 | 0.47 | 55 | 54 | 109 |
| 3 | -$22 (0.6) | 0.5 | 0.37 | 0.71 | 30 | 89 | 119 |
| 4 | $8 (0.7) | 0.55 | 0.96 | 0.11 | 62 | 8 | 70 |
| 5 | -$28 (0.55) | 0.6 | 0.97 | 0.08 | 102 | 4 | 106 |
| 6 | $8 (0.7) | 0.55 | 0.96 | 0.11 | 62 | 8 | 70 |

Table 2. Best performance thresholds and outcomes for each model.

## Discussion

This study demonstrates the potential of an impactability-based framework for resource allocation, one that has been enabled by harnessing the combined power of digital health, machine learning models and large datasets. We took an approach that first predicted future costs without a digital health intervention and compared predicted and observed costs. An impactable population was defined where these two figures deviated. Next, a further machine learning model was trained on data from the impactable population to subsequently be able to accurately predict a patient's ability to benefit (incur lower healthcare costs) if onboarded onto the digital health solution. Grid search was then performed to identify the optimal thresholds for impactability and confidence in prediction thereof. Far from being a definitive methodology, this study sought to demonstrate the utility of measuring impactability and advance a general framework for doing so with a roadmap for how both methodology and performance could be improved.

Current 'risk score' approaches assume every patient is potentially impactable and therefore fail to take into account the many factors that can contribute to a patient's likelihood of being affected by care.[24] Previous work by DuBard and Jackson[25] found only a 53% overlap between the patients with the highest impactability scores and those with the highest risk of inpatient admission, concluding in this instance that traditional risk-based methods are missing vital cost saving opportunities. Impactability is agnostic to which patients are most at risk of a certain outcome; the work described here defines it as attaining a threshold of cost saving as a consequence of onboarding onto a digital health intervention. Furthermore, a risk score (primarily derived from an amalgam of different parts of a patient's disease profile) is invalidated by virtue of not being modifiable by a patient's potential impactability with a given intervention.

Going against the convention of targeting those who currently have high cost or utilization -- or those with the highest predicted risk thereof -- justifiably raises the question of fairness and concerns around algorithmic bias, a particular concern with machine learning models.[25] However, the startup and running costs of the digital health tools at the core of this study is low, and therefore wide, inclusive rollout is made possible. The quantity of saving at an individual level (at times in the tens of dollars) can appear low, but the sum of potential savings is considerable when considering the much lower cost of the digital health intervention and the large populations of patients onboarded. It is within the significant non-operational budgets, such as marketing, where impactability can offer a targeted approach that channels resources towards recruiting those most likely to achieve the reduced costs offered by the intervention. For non-digital health interventions this framework can be used to determine how to best allocate care resources in general.

Both the impactable and non-impactable population in this study are smartphone users. It has been suspected that smartphone use is a proxy for other propitious social factors however it is plausible that as smartphones near ubiquity this 'digital divide' will narrow and this association will weaken.[26] In sum, the generalisability of this impactability model to populations of non-smartphone users is not clear. The methodology can be used to examine the impactability of non-digital healthcare interventions and this is the subject of further investigation by the research team at Wellframe. Further work is in progress by a research group in North Carolina that leverages an impactability score for allocation of pregnancy care management by identifying those patients at-risk of low-weight births whose characteristics match those who have previously benefited from additional care.[27]

As frameworks centered around impactability mature and access and use of digital health platforms increases, a new opportunity will be enabled: creating a closed-loop feedback system that lends real-time insights into a patient's ability to benefit from an intervention. This allows not only propitious selection but per-patient customisation, moving away from one-size-fits all care models in favour of data-driven individualised and responsive care.

While promising for use in future intervention targeting, this work does have limitations. Perhaps the biggest hindrance to model performance is the relatively small sample size due to the fact that the data was collected from an initial rollout phase of the digital health solution in question. The sample size constraint resulted in small training and test sets, and prohibited the creation and use of a validation set. As more patients onboard, more data will be available to train both the cost prediction model and the impactability model and a validation set will be used, resulting in better predictions of cost and impactability status. Our models also suffer from aggregated demographic variables since the only demographic variable in the dataset was 5 digit zip code. Knowing a patient's individual education level, income, employment status, and poverty status would result in more accurate predictions. Moreover, other demographic variables such as number of individuals in a patient's household and race would add to the predictive power of the models. The zip code level variables we use act as proxies for individual demographics are less reliable and more imprecise. Another limitation is the amount of follow-up time for each individual. Most patients in this population had approximately 12 months of claims prior to onboarding. Having more historical data may elucidate temporal patterns in care and improve future cost prediction and impactability classification. A matched control group from the same population was not used for comparison due to strong selection bias for patients who used the digital health intervention. Patients who were recruited and chose to participate saw a markedly different cost pattern in the months immediately prior to those who were recruited but chose not to participate. This fundamental difference in populations was not able to be controlled for by propensity score matching. This analysis is currently limited to one payer's commercially insured population. To establish the external validity of this model it will be tested on other payer's populations and ultimately across payer's. Additional data from other payers is likely to improve generalisability and increase model performance. It may also highlight differences in patient populations that may be beneficial to prediction and classification.

## Conclusion

With Digital Health technologies, health plans have the opportunity to deliver preventative interventions against incipient chronic disease that encompass more than just the highest risk populations. Machine learning models' definition of impactability assists in effective spending on marketing these interventions to patients. Though the concept of assigning impactability scores may seem like an endpoint, such predictive analytics are only set to improve, offering continually iterating insights that are responsive to change and help tailor interventions that enable cost-effective use of valuable resources. Digital Health has reduced the cost of care delivery such that much larger populations can be served in a cost effective manner hitherto impossible with a visit or telephone based care model. With analysis of impactability as described in this paper, it is possible to create a closed loop feedback system for continuous improvement in healthcare where informed patient selection is married with digital health interventions dynamically adapted to minimize the delta between observed and expected benefit, with the data generated used to further improve patient targeting. This work is a proposal for a general approach to patient targeting based on impactability instantiated in a specific model with a roadmap to improve model performance and generalizability.